\documentstyle[11pt,newpasp,twoside]{article}
\def\edcomment#1{\iffalse\marginpar{\raggedright\sl#1\/}\else\relax\fi}
\reversemarginpar

\begin{document}
\title{Intergalactic Globular Clusters}
\author{Michael J. West}
\affil{Department of Physics \& Astronomy, University of Hawaii, Hilo, Hawaii 96720, USA}
\author{Patrick C\^ot\'e}
\affil{Department of Physics \& Astronomy, Rutgers University, Piscataway, NJ 08854, USA}
\author{Henry C. Ferguson}
\affil{Space Telescope Science Institute, 3700 San Martin Drive, Baltimore, MD 21218, USA}
\author{Michael D. Gregg}
\affil{Department of Physics, University of California, Davis, CA 95616, USA; and
IGPP, Lawrence Livermore National Laboratory, Livermore, CA 94550, USA}
\author{Andr\'es Jord\'an}
\affil{Department of Physics \& Astronomy, Rutgers University, Piscataway, NJ 08854, USA}
\author{Ronald O. Marzke}
\affil{Department of Physics \& Astronomy, San Francisco State University, San Francisco, CA 94132, USA}
\author{Nial R. Tanvir}
\affil{Department of Physical Science, University of Hertfordshire, Hatfield AL10 9AB, UK}
\author{Ted von Hippel}
\affil{Department of Astronomy, University of Texas, Austin, TX 78712, USA}

\begin{abstract}
We confirm and extend our previous detection of a population of 
intergalactic globular clusters in Abell 1185, and 
report the first discovery of an intergalactic globular cluster in the nearby Virgo cluster of galaxies. 
The numbers, colors and luminosities of these objects can place constraints on their origin, 
which in turn may yield new insights to the evolution of galaxies in dense environments.
\end{abstract}

\section{Introduction}
There are several reasons to believe that a population of intergalactic globular clusters (IGCs) 
should exist outside of galaxies:

(1) The Jeans mass at recombination  
was $\sim 10^5 - 10^6$ solar masses, and hence globular cluster 
sized objects could have formed  
wherever the local density of matter was high enough.

(2) Many galaxies may have met their demise over a Hubble time
as a result of collisions and tidal disruption.  
Globular clusters  
are likely to survive the disruption
of their parent galaxy, resulting in the gradual accumulation of a population of IGCs.
Intergalactic stars, 
planetary nebulae, supernovae and 
HII regions have already been found; it would be surprising if there were no IGCs.

(3) The existence of IGCs might explain high specific frequencies, 
bimodal globular cluster metallicity distributions and other current puzzles in 
the study of 
%extragalactic 
globular cluster systems.

Jord\'an et al. (2003) reported a tentative detection of IGCs  
in the center of the rich galaxy cluster A1185 ($z = 0.032$) based on $I$-band images obtained 
with WFPC2 on the Hubble Space Telescope.   

\section{What's New?}
We (C\^ot\'e, Jord\'an, Marzke, West) recently obtained very deep, multicolored ($V$ and $I$) 
images of the 
same A1185 field using HST with the new ACS.  The goals of these new observations are to 1) 
detect the peak of the assumed universal Gaussian-like globular cluster luminosity function (which 
should occur at $I \sim 27.3$ at  A1185's distance) and thereby confirm that these candidate 
IGCs are bona fide globular clusters and 
2) use color information to infer their metallicities.  Preliminary analysis indicates that 
we are reaching sufficiently faint magnitudes to reliably detect the luminosity function turnover.
The number and colors (metallicities) 
of IGCs will provide constraints on the number and types of galaxies that have been 
destroyed or stripped over a Hubble time.

Using the Keck telescope, we (Ferguson, Gregg, Tanvir, von Hippel, West) recently measured the  
redshift of a candidate IGC in the nearby Virgo galaxy cluster  
that was found serendipitously on an HST image obtained for 
another project.   Preliminary data reductions show that this object, which is slightly resolved in the HST 
image and appears to be a distant globular cluster, 
has a recessional velocity of $\sim 470$ km/s, and hence is most likely in the Virgo cluster.  
Its apparent magnitude, $m_{V} \sim 21.2$, is consistent with it being a bright globular cluster.  
Using telescopes on Mauna Kea we have since obtained optical and NIR colors 
of this object, as well as a medium-resolution spectrum that should yield its velocity dispersion.  
These data are presently being analyzed.     

\acknowledgments MJW acknowledges support from NSF grant AST 02-05960 and HST grant HST-GO-09438.01-A.

\end{document}